\documentclass[12pt,preprint]{aastex}

\shorttitle{$^{6}$Li/$^{7}$Li in stars with planets}
\shortauthors{Ghezzi et al.}

\begin{document}


\title{Measurements of the Isotopic Ratio $^{6}$Li/$^{7}$Li in Stars with Planets}


\author{L. Ghezzi\altaffilmark{1}, K. Cunha\altaffilmark{2,1}, V. V. Smith\altaffilmark{2}, S. Margheim\altaffilmark{3}, S. Schuler\altaffilmark{2}, F. X. de Ara\'ujo\altaffilmark{1} and R. de la Reza\altaffilmark{1}}


\altaffiltext{1}{Observat\'orio Nacional, Rua General Jos\'e Cristino, 77, 20921-400, 
                 S\~ao Crist\'ov\~ao, Rio de Janeiro, RJ, Brazil; luan@on.br}
\altaffiltext{2}{National Optical Astronomy Observatory, 950 North Cherry Avenue, Tucson, AZ 85719, USA}
\altaffiltext{3}{Gemini Observatory, Casilla 603, La Serena, Chile}


\begin{abstract}

High-resolution (R = 143,000), high signal-to-noise (S/N = 700--1100)
Gemini-S bHROS spectra have been analyzed in a search for $^{6}$Li in 5 stars
which host extrasolar planets.  The presence of detectable amounts
of $^{6}$Li in these mature, solar-type stars is a good monitor of
accretion of planetary disk material, or solid bodies themselves,
into the outer layers of the parent stars.  Detailed profile-fitting
of the \ion{Li}{1} resonance doublet at $\lambda$6707.8 \AA\ reveals
no detectable amounts of $^{6}$Li in any star in our sample. The
list of stars analyzed includes HD 82943 for which $^{6}$Li has been previouly 
detected at the level of $^{6}$Li/$^{7}$Li = 0.05 $\pm$ 0.02. 
The typical limits in the derived isotopic fraction are 
$^{6}$Li/$^{7}$Li $\le$ 0.00--0.02. These upper limits 
constrain the amount of accreted material to less than $\sim$ 0.02 to
0.5 Jovian masses.  The presence of detectable amounts of $^{6}$Li
would manifest itself as a red asymmetry in the \ion{Li}{1} line-profile and
the derived upper limits on such asymmetries are discussed
in light of three-dimensional hydrodynamic model atmospheres, where
convective motions also give rise to slight red asymmetries in
line profiles.

\end{abstract}

%% example has been keyed in \apj style. See the instructions to authors




\section{INTRODUCTION}

\label{int}

One of the interesting properties of the known stars with planets concerns their metallicity distribution.
Several studies (\citealt{s00}; \citealt{g01}; \citealt{l03}; \citealt{s05}; \citealt{fv05}) have
confirmed the result first shown by \citet{g97}: stars with giant planets are
systematically metal-rich (by $\sim$ 0.2 dex) relative to field FGK dwarfs not known to
harbor planets. Two hypotheses have been proposed to explain this excess:
primordial enrichment or pollution. The first indicates that the probability of forming giant planets is a steeply
rising function of the intrinsic metallicity of the gas and dust cloud which gave birth to the system.
This possibility is in agreement with the core-accretion scenario (e.g., \citealt{p96}). The higher metal
content would raise the surface density of solid material in the disk, leading to a more efficient
agglutination of the cores onto which the gas will be accreted. The pollution scenario, on the
other hand, indicates that the presence of planets could alter the metallicity of their hosting stars. During
the inward migration process of giant planets, solid material from the protoplanetary disk or even
inner planetesimals and planets could be accreted into the convective envelope of the hosting star. As this
material is depleted in H and He, the star's metallicity would be enhanced.

Current results (e.g., \citealt{fv05}) show that the frequency of planets increases significantly for
higher metallicities, thus giving strong support for the primordial hypothesis. Evidence
for the occurence of pollution is still ambiguous. For instance, \citet{e06} studied the relation between
chemical abundances of several elements and their respective condensation temperatures in stars with and without planets, 
finding no signifcant differences in the two groups. 
On the other hand, \citet{p07} analyzed the metallicity distributions of planet-hosting dwarfs and giants and found that the
latter do not favor metal-rich systems. The authors argue that this result could be a strong indication
of pollution, as the metal excess could be erased by the dilution process that takes place during the later
stages of stellar evolution.

Another ambiguous result is the possible detection of $^{6}$Li in the atmospheres of stars with planets,
which is a sensitive test of the pollution hypothesis. 
Both lithium isotopes are destroyed at relatively low temperatures ($T=2 \times 10^{6}$ K for 
$^{6}$Li and $T=2.5 \times 10^{6}$ K for $^{7}$Li)
in stellar interiors. During the early stages of evolution in solar-type stars (specifically,
before entering the main-sequence), these stars are entirely convective and most of the primordial
Li is transported to deeper and hotter layers, where it is rapidly burned. The fraction of
lithium destruction is, however, a strong function of the stellar mass. For a given metallicity, there
is a mass range in which the $^{6}$Li is completely destroyed, while a significant amount of
$^{7}$Li is preserved. The lower edge of this range falls just above the solar mass
(corresponding to main-sequence late-F spectral types) for near solar metallicities. Thus, one should
not expect to find any $^{6}$Li in the atmospheres of solar-type stars. Any positive detection
could indicate an external contamination or pollution process.

\citet{i01} measured the isotopic ratio $^{6}$Li/$^{7}$Li in HD 82943 (with
two close-in giant planets) and found $^{6}$Li/$^{7}$Li = 0.126 $\pm$ 0.014.
This can be compared to a solar system (meteoritic) ratio of  $^{6}$Li/$^{7}$Li = 0.08.
The positive $^{6}$Li detection for this star was interpreted as observational
evidence of the pollution process. \citet{r02} studied 
$^{6}$Li in 8 planet hosting stars (HD 82943 included)
and found no significant amount of this isotope in HD 82943, nor in any of the targets analyzed. 
The difference from the previously published $^{6}$Li detection for HD
82943 was attributed to the use of a more complete line list, although it noted the presence of
an unidentified absorption in the Li region.
\citet{i03} investigated the nature of the unindentified absorption feature at 6708.025 \AA, which
affects the Li I line, by
observing several stars of different effective temperatures. They concluded that a high excitation
\ion{Si}{1} line first proposed by \citet{m75} is more adequate than the \ion{Ti}{1} line used by \citet{r02}.
Adopting a revised line list and higher quality spectra, the authors performed a new
analysis of HD 82943 and measured $^{6}$Li/$^{7}$Li = 0.05
$\pm$ 0.02. Thus, this most recent result for HD 82943 gave additional support to the pollution scenario. 

More recently, \citet{m04}
made a major extension of the previous line lists (especially for the CN contribution) and
tested three different possibilities for the unidentified feature at 6708.025 \AA\ (\ion{Si}{1},
\ion{Ti}{1} and \ion{Ti}{2}): for all of the 3 different line lists no $^{6}$Li was detected in
a sample of three planet-hosting stars. These
results are generally consistent with \citet{r02} and argue against the pollution scenario.
Unfortunatelly, HD 82943 was not analyzed by Mandell et al. (2004).

In this paper, the $^{6}$Li/$^{7}$Li isotopic ratio in HD 82943 is analyzed, as well as
in other four planet hosting stars and one star not known to have giant planets. 
The observations and the data reduction are described in \S \ref{obsred}. The analysis 
procedures including the derivation of atmospheric and broadening parameters and compilation of the
line list for the \ion{Li}{1} region are presented in  \S \ref{analysis}. The abundance results are
presented in \S \ref{li} and discussed in \S \ref{disc}.

%__________________________________________________________________

\section{OBSERVATIONS AND DATA REDUCTION}

\label{obsred}

\subsection{Observations}

\label{obs}

Spectra of the program stars were obtained at the Gemini-S telescope with the bench-mounted 
High-Resolution Optical Spectrograph (bHROS).  
Given the brightness of these targets, the \textquotedblleft object-only\textquotedblright\ mode was used for 
observations; for this mode, a 0.9\arcsec\ fiber is fed into an image slicer that produces a 
\textquoteleft slit\textquoteright, measuring 0.14\arcsec\ X 6.5\arcsec\ as projected to the camera 
focal plane.  The spectrograph is cross dispersed by a set of fused silica prisms
and an image slicer rotation mechanism is used to produce a \textquoteleft vertical\textquoteright\ 
slit at the observed central wavelength on the detector, a single 2048 X 4608 E2V CCD with 13.5 $\mu$m pixels.

The instrument was configured to produce a central wavelength on the detector 
of 6501 \AA. In this configuration, eight incomplete spectral orders were obtained, covering the 
interval between 5580--7230 \AA.  The continuous coverage available within a single echelle order varies 
from $\sim$ 50--70 \AA\ depending on the order.  The detector was used with 1x1 binning to achieve a 
resolution close to the spectrograph's nominal resolution of $R = \lambda/\delta\lambda$ = 
150,000 (3 pixel sampling) and minimize the impact of cosmic rays. The actual resolution was
measured using 10 ThAr lines in the order containing the \ion{Li}{1} feature and this resolution
was found to be R = 143,000 $\pm$ 5000. The spectra were obtained 
in a wide range of 
observing conditions during both classical (2006 May) and queue operations (2006 December and 
2007 January). Table \ref{obslog} contains a detailed log of the observations including spectral
types, V magnitudes, number of exposures, integration times,
and the resulting signal-to-noise ratios (per resolution element). The quoted S/N values are 
based on direct measurements of the rms in sample continuum regions which were selected 
based on inspection of the Solar Atlas. The measured values of signal-to-noise 
are compatible with the S/N based on Poisson statistics.

In addition to the target spectra, calibration sets 
of biases, flats, and ThAr arcs were obtained each night. When the observations were done in 
queue mode, only a single ThAr spectrum was taken each night as only a single object was observed 
per night. When observations were done in classical mode, ThAr spectra were taken which 
bracketed the observations. For example, the observations of HD 82943 obtained on May 9 2006 were bracketed 
with ThAr spectra and this provides a measure of the spectrograph stability. A direct measurement 
of the stability can be found when combining multiple spectra via crosscorrelation.
The largest drift found was about 3 m\AA\ or 1/5 of a pixel.
Telluric spectra were also obtained and, as expected, demonstrated no significant atmospheric 
contamination in the region of the lithium feature.

\subsection{Reduction}

\label{red}

The data were processed using standard echelle reduction practices with IRAF\footnote{IRAF (Image Reduction 
and Analysis Facility) is distributed by the National Optical Astronomy Observatories (NOAO), which is 
operated by the Association of Universities for Research in Astronomy, Inc. (AURA) under cooperative 
agreement with the National Science Foundation (NSO).} packages, 
following the recipe developed by the bHROS science demonstration team\footnote{Available at http://www.gemini.edu/sciops/instruments/hros/hrosIndex.html}.
The raw spectra were corrected by overscan and bias-image subtraction.  The images were 
then flat-fielded to remove both pixel-to-pixel variations and small amounts of fringing 
present in the spectra; since the illumination pattern on the detector is nearly identical for the flat 
and target spectra, 
the fringe pattern can be very well removed.  
Scattered light was then sampled, fit, and subtracted from our images before extraction.

\begin{figure}
\plotone{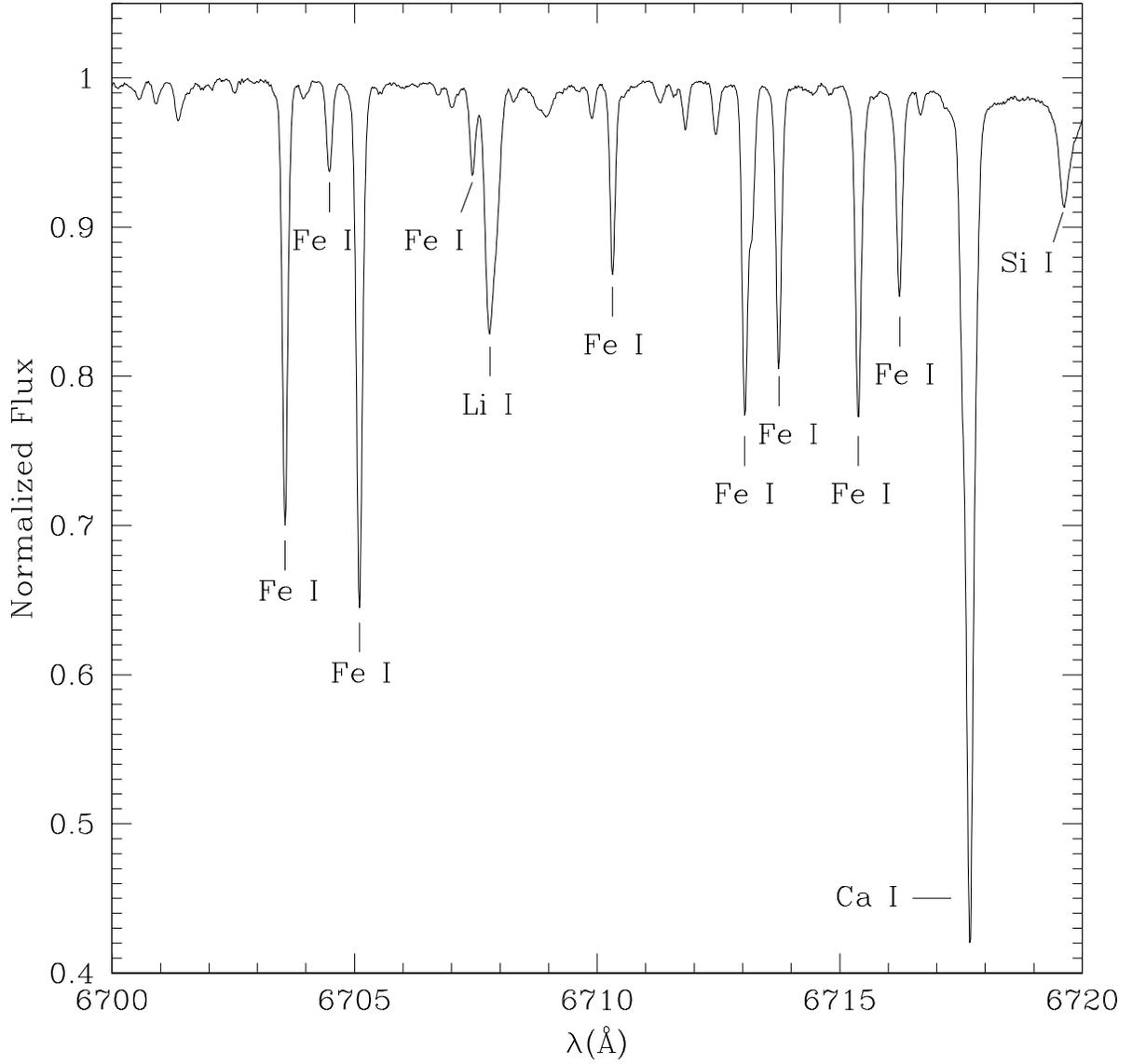}
\caption{The Gemini bHROS spectrum of HD 82943 (R $\simeq$ 143,000 and S/N $\simeq$ 1130), 
showing part of the spectral order which contains the lithium feature. The main
spectral lines in this region are identified.}
\label{spec}
\end{figure}

The use of the prism cross disperser combined with the long \textquoteleft 
slit\textquoteright\ length causes the spectral orders to become tilted away from the central order. 
The severity of this tilt can potentially result in the loss of resolution in the extraction 
process. To reduce the impact of the tilts, each of the eight spectral orders was divided 
into 14 subapertures, for a total of 112 resulting subapertures.  Each subaperture was individually 
extracted with optimal extraction method and blaze corrected with similarly extracted flat-field 
spectra. The ThAr spectra were also subdivided into the same subapertures and wavelength solutions 
were derived for each of the 112 subapertures, with a typical rms scatter of 0.001 \AA.  These 
wavelength solutions were applied to the target subapertures and combined to produce a reduced 
one-dimensional echelle spectrum composed of eight spectral orders. Rejection methods employed 
in the combination process ensured the removal of cosmic rays from the final spectra.
Finally, the spectra were continuum normalized and corrected for radial velocity shifts.
More details about the bHROS data reduction process can be found in \citet{s08}.
The final processed spectrum for a target star is shown in Figure \ref{spec}, as an example.

%%%%%%%%%%%%%%%%%%%%%%%%%%%%%%%%%%

\begin{deluxetable}{lcccccc}
\tablecolumns{9}
\tablewidth{0pc}
\tablecaption{Observing log.\label{obslog}}
\tablehead{
\colhead{Star} & \colhead{Spectral} & \colhead{$V$} & \colhead{Observation} & \colhead{N} & \colhead{$T_{\rm int}$} & \colhead{S/N} \\
\colhead{} & \colhead{Type} & \colhead{} & \colhead{Date} & \colhead{} & \colhead{(s)} & \colhead{($\sim$ 6708 \AA)}}
\startdata
HD 17051  & F9V  & 5.40 & 2006 Dec 13     & 1 & 600  & 740  \\
HD 36435  & G9V  & 7.01 & 2006 Dec 15     & 2 & 1050 & 560  \\
HD 74156  & G0   & 7.62 & 2007 Jan 30     & 3 & 1320 & 710  \\
HD 82943  & F9V  & 6.54 & 2006 May 09,10  & 4 & 600  & 1130 \\
HD 147513 & G1V  & 5.37 & 2006 May 10     & 4 & 600  & 1100 \\
HD 217107 & G8IV & 6.16 & 2006 May 10     & 3 & 600  & 750  \\
\enddata
\end{deluxetable}

%%%%%%%%%%%%%%%%%%%%%%%%%%%%%%%%%%%

%%%%%%%%%%%%%%%%%%%%%%%%%%%%%%%%%%%

%%%%%%%%%%%%%%%%%%%%%%%%%%%%%%%%%%%

%______________________________________________________________

\section{ANALYSIS}

\label{analysis}

The first step in the analysis is to derive effective temperatures, surface gravities, and
microturbulent velocities for the sample stars. In addition,
broadening parameters affecting the observed spectral lines also need to be defined.
A crucial point in the dertermination of $^{6}$Li/$^{7}$Li isotopic ratios
is the construction of a detailed line list for the region around the \ion{Li}{1} feature. 
This section presents a discussion of the analysis method adopted in this study.

\subsection{Line Selection and Derivation of Stellar Parameters}

\label{atmpar}

The bHROS spectra have incomplete wavelength coverage (Section \ref{obs})
which restricts the selection of iron lines to derive stellar parameters and metallicities. 
A sample of \ion{Fe}{1} and \ion{Fe}{2} lines was compiled from the list in \citet{t90} and 
equivalent widths were measured in the solar
spectrum \citep{k84} in order to select suitable lines and
\textit{gf}-values which produced an abundance scatter of less than 
or equal to 0.05 dex. The final list of adopted Fe lines, as well as
the measured equivalent widths for the target stars and the Sun are 
presented in Table \ref{ew}. The wavelenghts, lower excitation potentials (LEP),
and \textit{gf}-values were taken from the Vienna Atomic Line
Database\footnote{http://ams.astro.univie.ac.at/$\sim$vald/} (VALD-2; \citealt{k99}).
As a reference note the adopted line list yields a solar abundance 
A(Fe)\footnote {A(Fe)= log [N(Fe)/N(H)]+12}= 7.48 $\pm$ 0.05 and $\xi$ = 1.24 km $\rm s^{-1}$
using a Kurucz model atmosphere with $T_{\rm eff}$ = 5777 K, $\log g$ = 4.44,
$\xi$ = 2.00 km $\rm s^{-1}$, and l/H$_{p}$ = 1.25.

\begin{figure}
\plotone{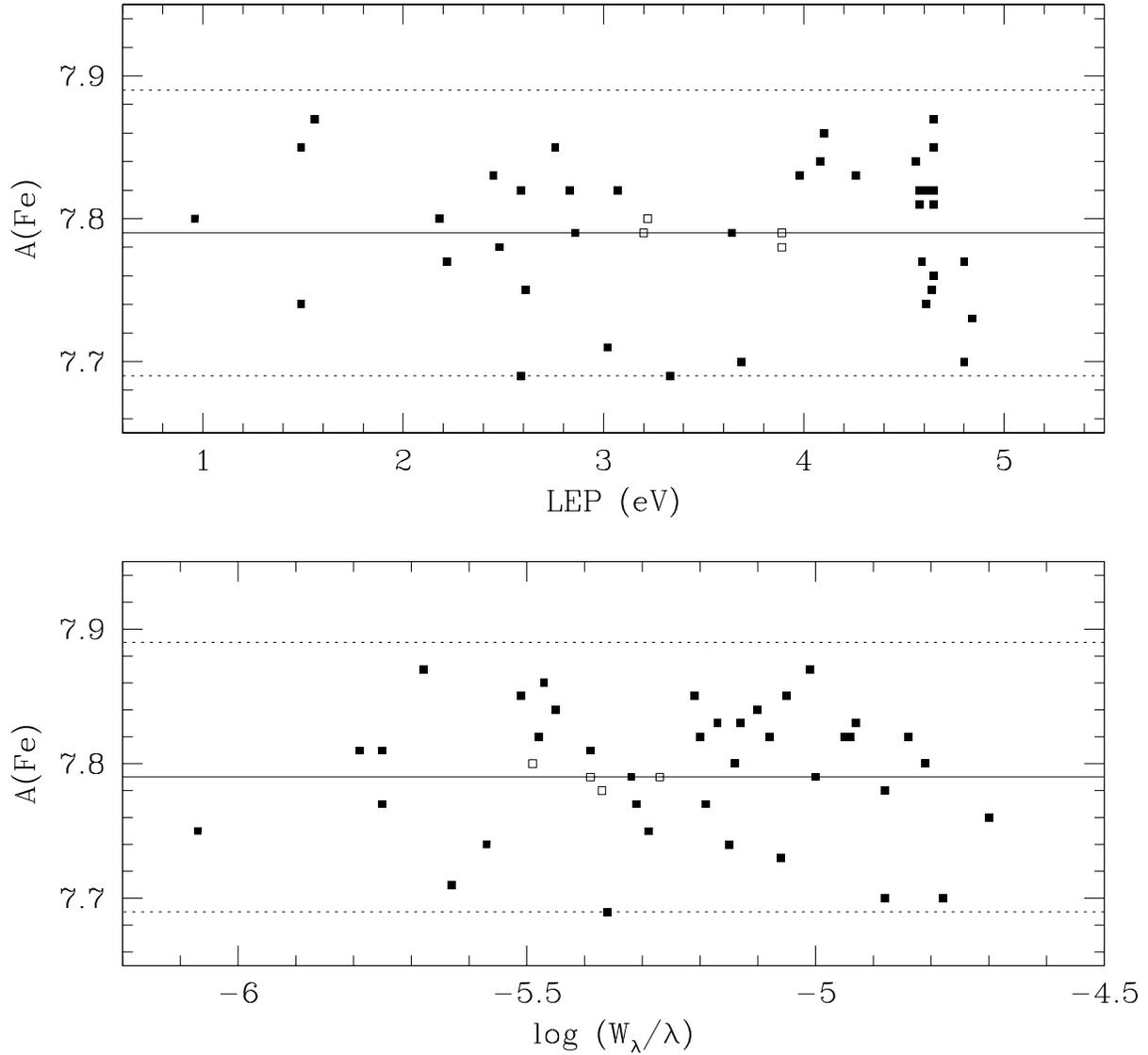}
\caption{Spectroscopic determination of the effective temperature and microturbulent velocity for HD 82943 
obtained from zero slopes in the runs of \ion{Fe}{1} abundances with excitation potential of the transitions (upper panel) and reduced equivalent widths (lower panel). \ion{Fe}{1} (filled squares) 
and \ion{Fe}{2} (open squares) abundances are consistent and 
the slopes are zero for A(Fe) = 7.79.}
\label{hd82943_atm}
\end{figure}

Stellar parameters for the target stars were derived spectroscopically
and followed standard techniques. Effective temperatures were
obtained from zero slopes in diagrams of Fe abundance versus excitation potential and
surface gravities from ionization equilibrium between \ion{Fe}{1} and \ion{Fe}{2} species.
The microturbulent velocities were varied until the slopes of
A(\ion{Fe}{1}) \textit{versus} $\log (W_{\lambda}/\lambda)$ were zero.
The abundances were derived in LTE using an updated
version of the spectrum synthesis code MOOG\footnote{Available at
http://verdi.as.utexas.edu/moog.html} \citep{s73}. The model atmospheres
adopted in the analysis were interpolated from the ODFNEW grid of ATLAS9 models\footnote{http://kurucz.harvard.edu/} \citep{ck04}.
Figure \ref{hd82943_atm} shows the iterated result of A(\ion{Fe}{1}) \textit{versus} LEP (top panel)
and A(\ion{Fe}{1}) \textit{versus} $\log (W_{\lambda}/\lambda)$ (bottom panel)
for the star HD 82943, as an example.
The adopted stellar parameters for all target stars are
given in Table \ref{atm}. Our adopted values for effective temperatures are in good 
agreement (within $\sim$100 K) with the derived values using the photometric 
correlations presented in \citet{rm04}. Differences of this amount have no significant
effect in the derived isotopic fractions.

The Fe abundance uncertainties can be estimated from the errors in $T_{\rm eff}$, $\log g$, 
$\xi$, line \textit{gf}-values, continuum placement, and equivalent width measurements, among others. 
We estimate the uncertainties in $T_{\rm eff}$ to be $\pm100$ K; log g $\pm$ 0.2 dex; and $\xi$ $\pm$ 0.2 km/s.
The combined uncertainties in the derived Fe abundances are typical for this type of classical
spectroscopic analysis and are estimated to be about $\pm$ 0.1 dex. 

The luminosities of the program stars were calculated using apparent
V-magnitudes, Hipparcos parallaxes, bolometric
corrections from Giradi et al. (2002), and using M$_{\rm bol,\odot}$ 
= 4.77.  No interstellar extinction was included as the most distant star
has d = 65 pc.  Utilizing these luminosities and the T$_{\rm eff}$ values 
from Table \ref{atm}, stellar masses were derived by placing the stars in a grid 
of evolutionary tracks from Girardi et al. (2000) for [Fe/H] = 0.00 and
0.20, with stellar masses covering M = 0.8 to 1.4 M$_{\odot}$.  Stellar
ages were estimated by using the values of effective temperature and
luminosity to place the stars in a grid of isochrones from Girardi et al.
(2000) for [Fe/H] = 0.00 and 0.20, with ages from 0.063 to 7 Gyr.  In 
addition, stellar masses were calculated using the relation between
luminosity, surface gravity, and effective temperature:
\begin{equation}
\label{mstar}
\log \frac{M}{M_{\odot}} = \log \frac{L}{L_{\odot}} + \log \frac{g}{g_{\odot}} - 4\log \frac{T_{\rm eff}}{T_{\rm eff,\odot}} ,
\end{equation}
with T$_{\rm eff,\odot}$ 
= 5777 K and log g$_{\odot}$ = 4.44. The masses and ages estimated for
the targets stars are presented in Table \ref{atm}.

%%%%%%%%%%%%%%%%%%%%%%%%%%%%%%%%%%

\begin{deluxetable}{ccccccccccc}
\tabletypesize{\scriptsize}
\tablecolumns{11}
\tablewidth{0pc}
\tablecaption{Selected Fe lines and Measured Equivalent Widths.\label{ew}}
\tablehead{
\colhead{$\lambda$} & \colhead{Ion} & \colhead{LEP} & \colhead{$\log gf$} & \multicolumn{7}{c}{W$_{\lambda}$ (m\AA)} \\
\colhead{(\AA)} & \colhead{} & \colhead{(eV)} & \colhead{(dex)} & \colhead{Sun} & \colhead{HD 17051} & \colhead{HD 36435} & \colhead{HD 74156} & \colhead{HD 82943} & \colhead{HD 147513} & \colhead{HD 217107}}
\startdata
5741.848 & \ion{Fe}{1}  & 4.26 & $-$1.730 & 32.9  & 35.4      & 41.9      & 30.8      & 38.8      & 32.1      & 52.7      \\
5760.345 & \ion{Fe}{1}  & 3.64 & $-$2.490 & 23.8  & 23.5      & 34.9      & 23.4      & 27.6      & 22.7      & 42.0      \\
5778.453 & \ion{Fe}{1}  & 2.59 & $-$3.430 & 22.7  & 20.9      & 35.6      & 20.7      & 25.0      & 20.8      & 45.5      \\
5905.672 & \ion{Fe}{1}  & 4.65 & $-$0.863 & 57.9  & 59.9      & 67.9      & 58.3      & 67.3      & 58.1      & 76.7      \\
5916.247 & \ion{Fe}{1}  & 2.45 & $-$2.832 & 62.2  & 62.9      & 77.6      & 58.5      & 70.0      & 62.7      & 86.2      \\
5927.789 & \ion{Fe}{1}  & 4.65 & $-$1.090 & 43.1  & \nodata   & 52.0      & 42.1      & 57.7      & 42.7      & 67.3      \\
5930.180 & \ion{Fe}{1}  & 4.65 & $-$0.028 & 105.7 & 112.1     & 129.4     & 103.2     & 119.4     & 107.7     & 136.1     \\
6079.009 & \ion{Fe}{1}  & 4.65 & $-$1.120 & 46.2  & 48.3      & 57.6      & 46.0      & 54.7      & 46.8      & 64.3      \\
6082.711 & \ion{Fe}{1}  & 2.22 & $-$3.573 & 34.8  & 28.8      & 52.1      & 29.8      & 39.4      & 33.5      & 60.9      \\
6093.644 & \ion{Fe}{1}  & 4.61 & $-$1.252 & 31.2  & 33.7      & 40.5      & 34.5      & 43.6      & 31.2      & 54.7      \\
6094.374 & \ion{Fe}{1}  & 4.65 & $-$1.661 & 20.2  & 21.7      & 25.3      & 20.2      & 25.1      & 19.5      & 36.9      \\
6096.665 & \ion{Fe}{1}  & 3.98 & $-$1.890 & 38.3  & 37.8      & 49.9      & 36.8      & 44.9      & 38.3      & 58.4      \\
6098.245 & \ion{Fe}{1}  & 4.56 & $-$1.859 & 16.2  & 18.8      & 23.3      & 16.8      & 21.8      & 16.3      & 33.0      \\
6265.134 & \ion{Fe}{1}  & 2.18 & $-$2.550 & 94.0  & 86.0      & 118.6     & 89.4      & 97.6      & \nodata   & \nodata   \\
6270.225 & \ion{Fe}{1}  & 2.86 & $-$2.576 & 57.6  & 48.7      & 68.0      & 54.8      & 62.2      & 54.6      & 78.0      \\
6271.279 & \ion{Fe}{1}  & 3.33 & $-$2.703 & 23.3  & 23.4      & 34.5      & 21.2      & 27.6      & 22.1      & 42.5      \\
6297.793 & \ion{Fe}{1}  & 2.22 & $-$2.740 & 81.0  & \nodata   & 104.0     & 78.2      & \nodata   & 85.1      & 105.7     \\
6302.494 & \ion{Fe}{1}  & 3.69 & $-$0.973 & 96.1  & 89.8      & \nodata   & 91.5      & 103.4     & 101.9     &     123.0 \\
6315.812 & \ion{Fe}{1}  & 4.08 & $-$1.710 & 41.7  & 41.9      & 52.3      & 43.0      & 50.4      & 42.4      & 62.6      \\
6322.686 & \ion{Fe}{1}  & 2.59 & $-$2.304 & 84.5  & 74.0      & 103.6     & 81.5      & 92.1      & 86.0      & 108.4     \\
6495.742 & \ion{Fe}{1}  & 4.84 & $-$0.801 & 48.4  & 45.0      & 58.9      & \nodata   & 57.2      & \nodata   & 72.4      \\
6496.467 & \ion{Fe}{1}  & 4.80 & $-$0.348 & 75.6  & 75.6      & 90.8      & 73.4      & 84.8      & 77.8      & 101.6     \\
6498.939 & \ion{Fe}{1}  & 0.96 & $-$4.699 & 45.6  & 34.8      & 72.6      & 39.8      & 47.2      & 42.3      & 76.5      \\
6518.367 & \ion{Fe}{1}  & 2.83 & $-$2.460 & 66.1  & 62.6      & 80.5      & 62.0      & 72.5      & 62.8      & 87.8      \\
6699.142 & \ion{Fe}{1}  & 4.59 & $-$2.101 & 9.1   & 9.2       & 11.9      & 8.2       & 12.0      & 8.0       & 18.0      \\
6703.567 & \ion{Fe}{1}  & 2.76 & $-$3.141 & 37.4  & 32.2      & 50.9      & 33.5      & 41.6      & 35.2      & 59.4      \\
6710.320 & \ion{Fe}{1}  & 1.49 & $-$4.764 & 16.1  & 11.6      & 27.3      & 12.3      & 18.0      & 13.7      & 35.4      \\
6713.745 & \ion{Fe}{1}  & 4.80 & $-$1.338 & 21.8  & 22.5      & 28.6      & 24.0      & 33.1      & 21.4      & 44.1      \\
6716.237 & \ion{Fe}{1}  & 4.58 & $-$1.836 & 16.0  & 17.3      & 22.6      & 16.6      & 22.1      & 15.3      & 31.6      \\
6725.357 & \ion{Fe}{1}  & 4.10 & $-$2.300 & 17.5  & 16.7      & 23.9      & 17.0      & 22.8      & 17.7      & 34.3      \\
6726.666 & \ion{Fe}{1}  & 4.61 & $-$1.133 & 46.6  & 47.6      & 59.8      & 45.8      & 56.0      & 46.6      & 66.6      \\
6732.065 & \ion{Fe}{1}  & 4.58 & $-$2.210 & 7.7   & 8.2       & 9.1       & 7.7       & 10.8      & 7.5       & 16.0      \\
6733.151 & \ion{Fe}{1}  & 4.64 & $-$1.437 & 26.4  & 26.8      & 32.7      & 25.9      & 34.5      & 25.5      & 48.6      \\
6739.522 & \ion{Fe}{1}  & 1.56 & $-$4.950 & 12.0  & 9.8       & 19.4      & 8.0       & 14.0      & 10.6      & 26.9      \\
6745.101 & \ion{Fe}{1}  & 4.58 & $-$2.160 & 9.0   & 10.6      & 11.3      & 8.9       & 11.9      & 8.3       & 18.1      \\
6746.955 & \ion{Fe}{1}  & 2.61 & $-$4.262 & 4.9   & 4.1       & 9.4       & 4.0       & 5.8       & 3.8       & 11.5      \\
6971.950 & \ion{Fe}{1}  & 3.02 & $-$3.340 & 13.8  & 11.7      & 21.1      & 12.5      & 16.5      & 12.3      & 29.4      \\
6978.852 & \ion{Fe}{1}  & 2.48 & $-$2.413 & 86.4  & 84.1      & 113.4     & 79.3      & 92.8      & 88.8      & 108.5     \\
7179.995 & \ion{Fe}{1}  & 1.49 & $-$4.780 & 20.4  & 14.9      & 33.0      & 14.6      & 22.3      & 19.2      & 42.4      \\
7189.146 & \ion{Fe}{1}  & 3.07 & $-$2.771 & 40.4  & 35.3      & 53.7      & 37.5      & 45.5      & 38.5      & 62.8      \\
6084.111 & \ion{Fe}{2}  & 3.20 & $-$3.780 & 22.0  & 32.2      & 18.5      & 31.7      & 32.8      & 21.6      & 31.0      \\
6113.322 & \ion{Fe}{2}  & 3.22 & $-$4.110 & 12.5  & 21.0      & 9.6       & 19.1      & 19.8      & 12.0      & 19.0      \\
7222.394 & \ion{Fe}{2}  & 3.89 & $-$3.276 & 19.7  & \nodata   & \nodata   & 29.6      & 29.1      & 19.2      & 27.1      \\
7224.487 & \ion{Fe}{2}  & 3.89 & $-$3.226 & 21.3  & 32.4      & 16.3      & 31.8      & 30.5      & 20.0      & \nodata   \\
\enddata
\end{deluxetable}

%%%%%%%%%%%%%%%%%%%%%%%%%%%%%%%%%%

%%%%%%%%%%%%%%%%%%%%%%%%%%%%%%%%%%%

%%%%%%%%%%%%%%%%%%%%%%%%%%%%%%%%%%

\begin{deluxetable}{lcccccc}
\tablecolumns{7}
\tablewidth{0pc}
\tablecaption{Adopted Stellar Parameters and Derived Metallicities.\label{atm}}
\tablehead{
\colhead{Star} & \colhead{$T_{\rm eff}$} & \colhead{$\log g$} & \colhead{$\xi$} & \colhead{M} & \colhead{Age} & \colhead{A(Fe)} \\
\colhead{} & \colhead{(K)} & \colhead{(cm $\rm s^{-2}$)} & \colhead{(km $\rm s^{-1}$)} & \colhead{(M$_{\odot}$)} & \colhead{(Gyr)}}
\startdata
HD 17051  & 6197 & 4.49 & 1.24 & 1.2 & 1.0 & 7.73 \\
HD 36435  & 5503 & 4.56 & 1.46 & 1.0 & 1.0 & 7.52 \\
HD 74156  & 6100 & 4.36 & 1.38 & 1.4 & 2.5 & 7.64 \\
HD 82943  & 6055 & 4.56 & 1.33 & 1.2 & 1.5 & 7.79 \\
HD 147513 & 5904 & 4.63 & 1.48 & 1.0 & 2.0 & 7.52 \\
HD 217107 & 5690 & 4.44 & 1.25 & 1.1 & 5.5 & 7.87 \\
\enddata
\end{deluxetable}

%%%%%%%%%%%%%%%%%%%%%%%%%%%%%%%%%%%

\subsection{Spectrum Synthesis}

\label{synth}

Lithium isotopic ratios are measured via spectrum synthesis analysis.
The \ion{Li}{1} doublet at $\sim$ 6708 \AA\ is an asymmetric blend of two
$^{7}$Li lines which consist of 7 hyperfine components, spanning a separation of
approximately 0.15 \AA\ between stronger blue and weaker red components. 
The $^{6}$Li lines present the exact same configuration, except with
3 hyperfine levels,  with a total separation
of approximately 0.15 \AA\ between stronger blue and weaker red components. As the $^{6}$Li
lines are much weaker, with the stronger one situated almost at the same wavelength as the weaker
$^{7}$Li component, the former isotope appears as a perturbation, providing an additional small
asymmetry to the blend. In addition, the lithium region suffers from the contribution of several blends
from metal and molecular CN lines. 

\subsubsection{Broadening Parameters}

\label{broadpar}

The stellar projected rotational velocity, $v$ sin \textit{i}, and
macroturbulent velocity, $V_{m}$, are broadening parameters which need 
to be defined for modelling spectral lines via spectrum synthesis. 
A good strategy to estimate rotation and macroturbulent velocities 
is to analyze lines which are isolated, unblended, and with similar 
strengths to the Li line
and then apply the same broadening parameters to the \ion{Li}{1} synthesis. 
A few \ion{Fe}{1} lines which fall in the same echelle order as the \ion{Li}{1} feature were investigated
and the \ion{Fe}{1}\ at 6703.567 \AA\ was identified as a clean line which was
used to estimate $v$ sin \textit{i} and $V_{m}$ for the studied stars.

A grid of synthetic spectra was computed for combinations of $v$ sin \textit{i} and $V_{m}$ varying between 
0 and 10 km $\rm s^{-1}$ (with steps of 0.1 km $\rm s^{-1}$). 
Also, we let the iron abundance vary within 0.05 dex of the abundance value which
was obtained for that line.
Best fits between 
synthetic and observed line profiles were obtained from a $\chi^{2}$ minimization 
as follows:
\begin{equation}
\label{chi2}
\chi_{r}^{2} = \frac{1}{(d-1)} \sum_{i=1}^{n} \frac{(O_{i}-S_{i})^2}{\sigma^2} ,
\end{equation}
where $O_{i}$ and $S_{i}$ are, respectively, the observed and synthetic 
normalized fluxes at a wavelength point \textit{i} across the line profile;
$\sigma$ is rms error of the continuum, given by $(\rm S/N)^{-1}$; $d=n-p$ is 
the number of degrees of freedom in the fit, where \textit{n} is the number of points in 
the observed spectra used in the fit and \textit{p} is the number of free parameters in 
the calculation of the synthetic spectra. In this case,
$p=5$: $v$ sin \textit{i} and $V_{m}$; \textit{r} (continuum level), \textit{w} (wavelength) 
and iron abundance A(Fe).
Small adjustments in the continuum level ($r \leq 0.4\%$),
to account for possible errors in the normalization process were allowed.
In addition, shifts in the central wavelenghts
of the \ion{Fe}{1} lines were needed in order to properly match the observed lines.
 
Since the resolution of the bHROS spectra is R $\sim$ 143,000, the
velocity resolution is just a bit larger than 2 km s$^{-1}$.  Thus,
at low values of $v$ sin \textit{i} and typical macroturblent velocities in stars
of the type studied here, the spectra will not be sensitive to changes
in low projected rotational velocities found in the program stars.
In order to estimate a lower limit to $v$ sin \textit{i} that can be detected
significantly with the data here, synthetic spectra were computed for
the \ion{Fe}{1} 6703.567 \AA\ line with a range of values of $v$ sin \textit{i}, from 1.0 to 3.0
km s$^{-1}$, a single macroturbulent velocity of 3.0 km s$^{-1}$
and additionally smoothed to the bHROS resolution, using a model with
solar parameters.  The synthetic spectra were then sampled at the same scale as bHROS, 
after which noise was added such that S/N = 1,000.  These
\textquotedblleft degraded\textquotedblright\ synthetic spectra were then subjected to the same analysis
as the real spectra, with the result that for $v$ sin \textit{i} values of 3.0 and
2.5 km s$^{-1}$ the proper rotational velocities were recovered to
within 0.3 km s$^{-1}$ (close to what the estimated uncertainties
are for the target stars).  For values of 2.0 and 1.5 km s$^{-1}$,
the analysis yielded a lower and constant value of 1.3 km s$^{-1}$,
indicating that the bHROS spectra are not necessarily sensitive to
values of $v$ sin \textit{i} $\le$ 1.5--2.0 km s$^{-1}$ for typical macroturbulent velocities.

We note, however, that even though the bHROS spectra are not sensitive to projected
rotational velocities of less than about 2.0 km s$^{-1}$, it is
still found in the real data that small values of $v$ sin \textit{i} provide
somewhat better fits to the \ion{Fe}{1} line profiles than a combination
of larger $v$ sin \textit{i}  and lower macrotubulence.  Although these lower values of
$v$ sin \textit{i}  provide better fits, their detection should not be considered
significant, with realistic lower limits of $v$ sin \textit{i} = 1.5 km/s.

\begin{figure}
\plotone{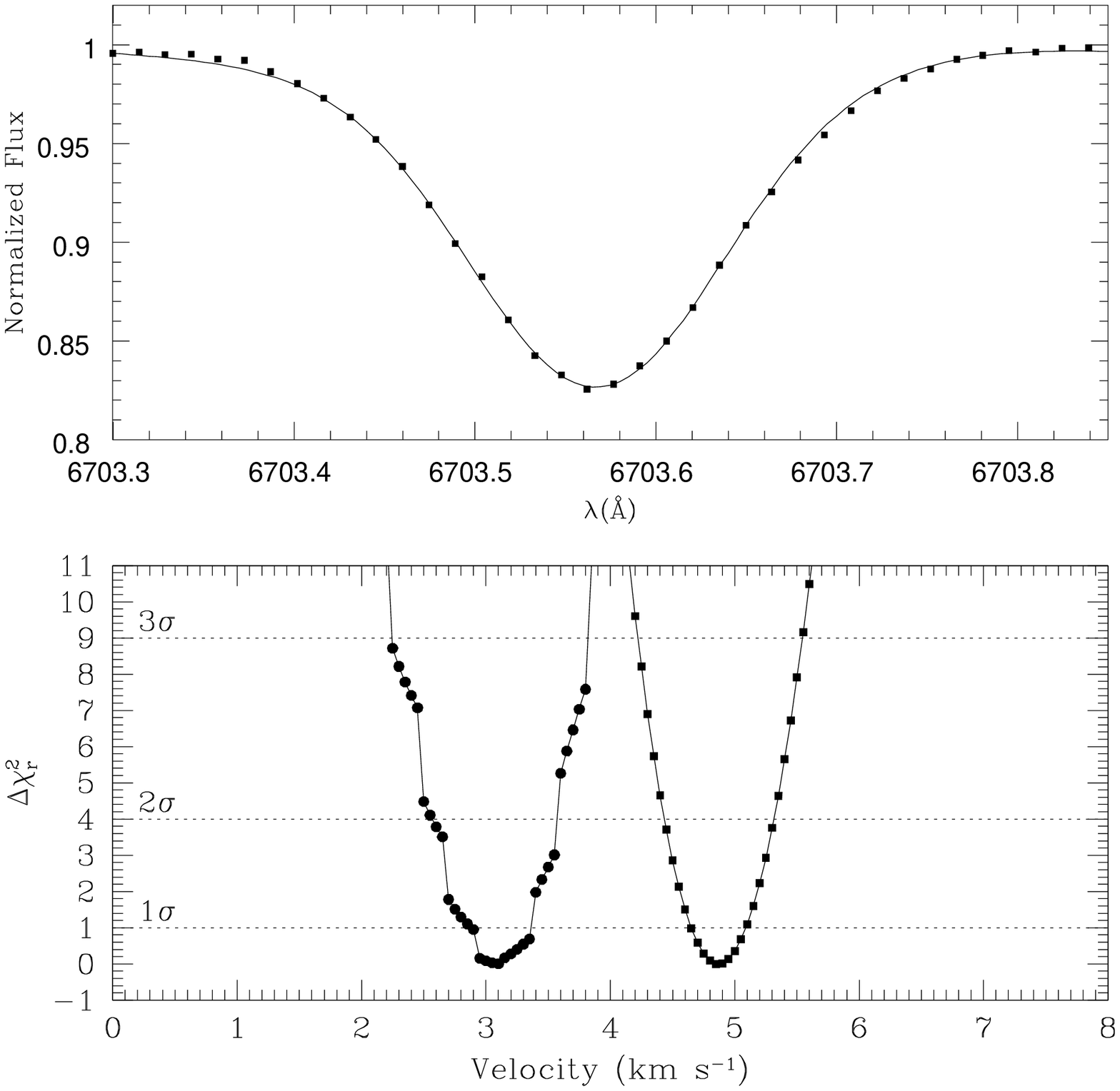}
\caption{Top Panel: Observed (filled squares) and synthetic (solid line) spectra
for the selected \ion{Fe}{1} line which was used to define 
the broadening parameters in target star HD 74156. 
The $\chi_{r}^{2}$-minimization used for estimating $v$ sin \textit{i} (filled circles; left)
and V$_{\rm m}$ (filled squares; right) is shown in the bottom panel.}
\label{hd74156_broad}
\end{figure}

Figure \ref{hd74156_broad} shows the fit obtained for the star HD 74156.  
The top panel plot both the observed and best-fit synthetic \ion{Fe}{1} line profile as
an illustration of the quality of the fits that are obtained in the
analysis.  
The bottom panel illustrates the 
$\chi^{2}$-minimization techniques used for estimating both projected rotational 
velocity (left) and macroturbulence (right).

The uncertainties in the overall line broadening caused by 
macroturbulent and projected rotational velocities
can be estimated by varying $v$ sin \textit{i} (while keeping V$_{\rm m}$ fixed)
in an interval of 1.0 km s$^{-1}$ around its best value, with steps of
0.05 km s$^{-1}$.  For each value of $v$ sin \textit{i}, $\Delta\chi_{\rm r}^{2}$
= $\chi_{\rm r}^{2}$ - $\chi_{\rm r,min}^{2}$ was calculated
and the velocity change which produced a 1$\sigma$ change in $\Delta\chi_{\rm r}^{2}$
was taken as the uncertainty.  
The uncertainties in V$_{\rm m}$ were obtained in a similar manner.
The adopted broadening parameters and uncertainties are presented in Table \ref{isotop}.

%%%%%%%%%%%%%%%%%%%%%%%%%%%%%%%%%%%

%%%%%%%%%%%%%%%%%%%%%%%%%%%%%%%%%%%

%%%%%%%%%%%%%%%%%%%%%%%%%%%%%%%%%%%

\begin{deluxetable}{lcccc}
\tablecolumns{5}
\tablewidth{0pc}
\tablecaption{Elemental abundances and isotopic ratios.\label{isotop}}
\tablehead{
\colhead{Star} & \colhead{$v$ sin \textit{i}} & \colhead{$V_{m}$} & \colhead{A(Li)} & \colhead{$^6$Li/$^7$Li} \\
\colhead{} & \colhead{(km s$^{-1}$)} & \colhead{(km s$^{-1}$)} & \colhead{} & \colhead{}} 
\startdata
Sun        &  1.70            & 2.80            & 0.96            & 0.00            \\
HD 17051   &  4.90 $\pm$ 0.25 & 5.25 $\pm$ 0.30 & 2.48 $\pm$ 0.01 & 0.03 $\pm$ 0.04 \\
HD 36435   &  4.25 $\pm$ 0.20 & 2.95 $\pm$ 0.30 & 1.60 $\pm$ 0.03 & 0.06 $\pm$ 0.08 \\
HD 74156   &  3.10 $\pm$ 0.25 & 4.85 $\pm$ 0.25 & 2.59 $\pm$ 0.01 & 0.00 $\pm$ 0.03 \\
HD 82943   &  $\le 1.5$       & 3.60 $\pm$ 0.10 & 2.49 $\pm$ 0.01 & 0.00 $\pm$ 0.02 \\
HD 147513  &  $\le 1.5$       & 3.05 $\pm$ 0.10 & 2.03 $\pm$ 0.01 & 0.02 $\pm$ 0.03 \\
HD 217107  &  $\le 1.5$       & 2.85 $\pm$ 0.10 & $\le 0.36$      & 0.00 $\pm$ 0.04 \\
\enddata
\end{deluxetable}

%%%%%%%%%%%%%%%%%%%%%%%%%%%%%%%%%%%

\subsubsection{Convection and line asymmetries}

\label{conv}

\begin{figure}
\plotone{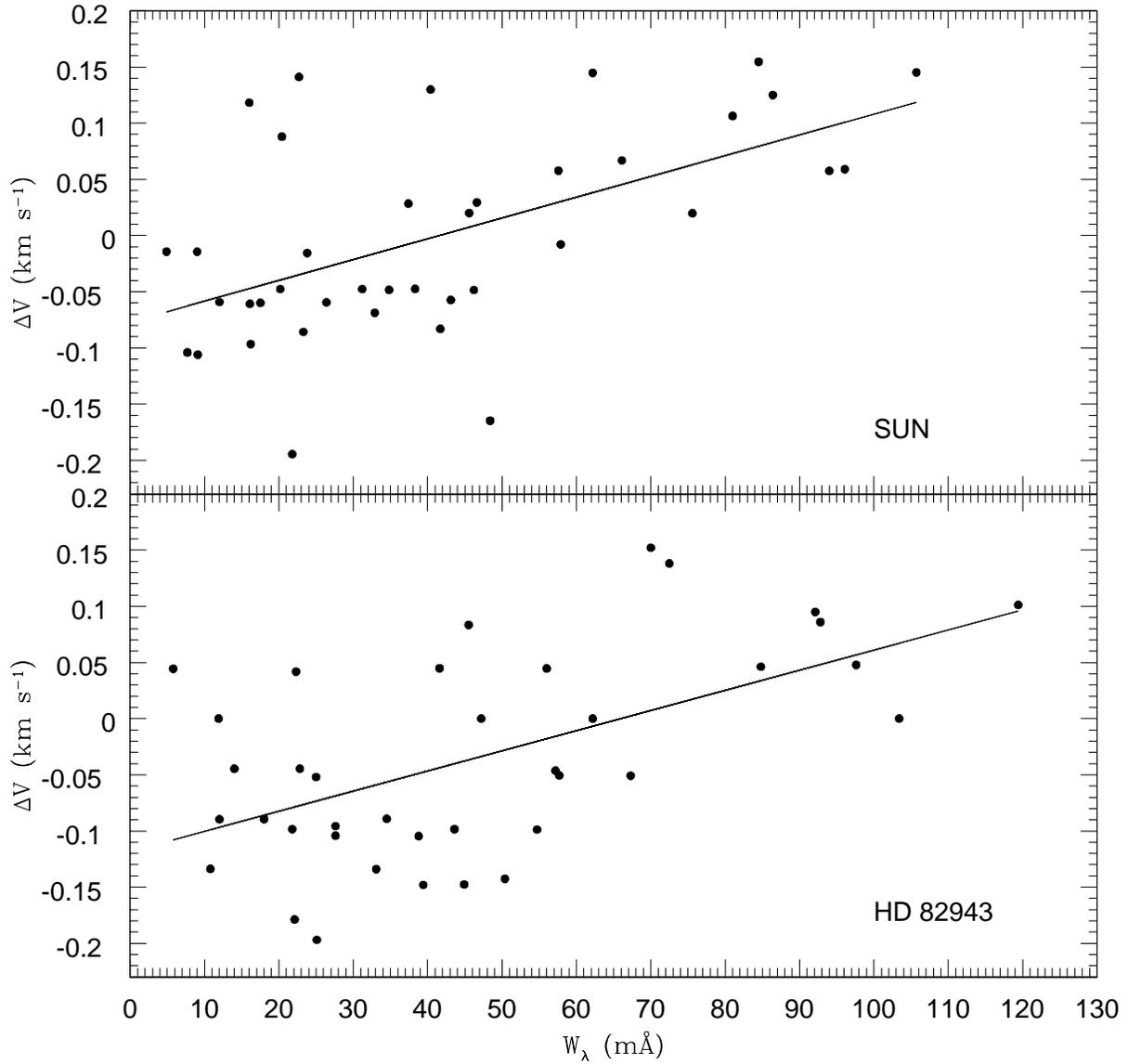}
\caption{Convective shifts measured for \ion{Fe}{1} lines in the solar spectrum
(top panel) and HD 82943 (bottom panel). The solid lines in each panel
represent the least-squares fits to the points.}
\label{convective_shifts}
\end{figure}

The additional small radial velocity shifts noted above represent convective shifts which are related
to the effects of granulation in the stellar atmospheres (see discussion in \citealt{ap02}). 
Figure \ref{convective_shifts} shows the trend of radial velocity shifts applied as a function of the 
measured equivalent widths for target star HD 82943 and the Sun.
The trend of increasingly positive radial velocities as the equivalent widths increase
has been noted previously by \citet{ap02}, \citet{r02}, and \citet{m04}. 
All three of these studies measured the Sun and fit straight lines
to the trends with slopes and intercepts respectively of: 2.9 m $\rm s^{-1}$ (m\AA)$^{-1}$ and 80 m $\rm s^{-1}$ \citep{ap02};
2.8 m $\rm s^{-1}$ (m\AA)$^{-1}$ and 120 m $\rm s^{-1}$ \citep{r02}; 4.2 m $\rm s^{-1}$ (m\AA)$^{-1}$ and 30 m $\rm s^{-1}$ \citep{m04}. Since different Fe lines were used by the different groups, as well as different
fitting techniques, the differences between the trends are not surprising and do not
lead to significant differences in convective shifts. The results reported here
for the Sun are a slope of 2.0 m $\rm s^{-1}$ (m\AA)$^{-1}$ and an intercept of $-$80 m $\rm s^{-1}$.

%%%%%%%%%%%%%%%%%%%%%%%%%%%%%%%%%%%

%%%%%%%%%%%%%%%%%%%%%%%%%%%%%%%%%%%%

\begin{figure}
\plotone{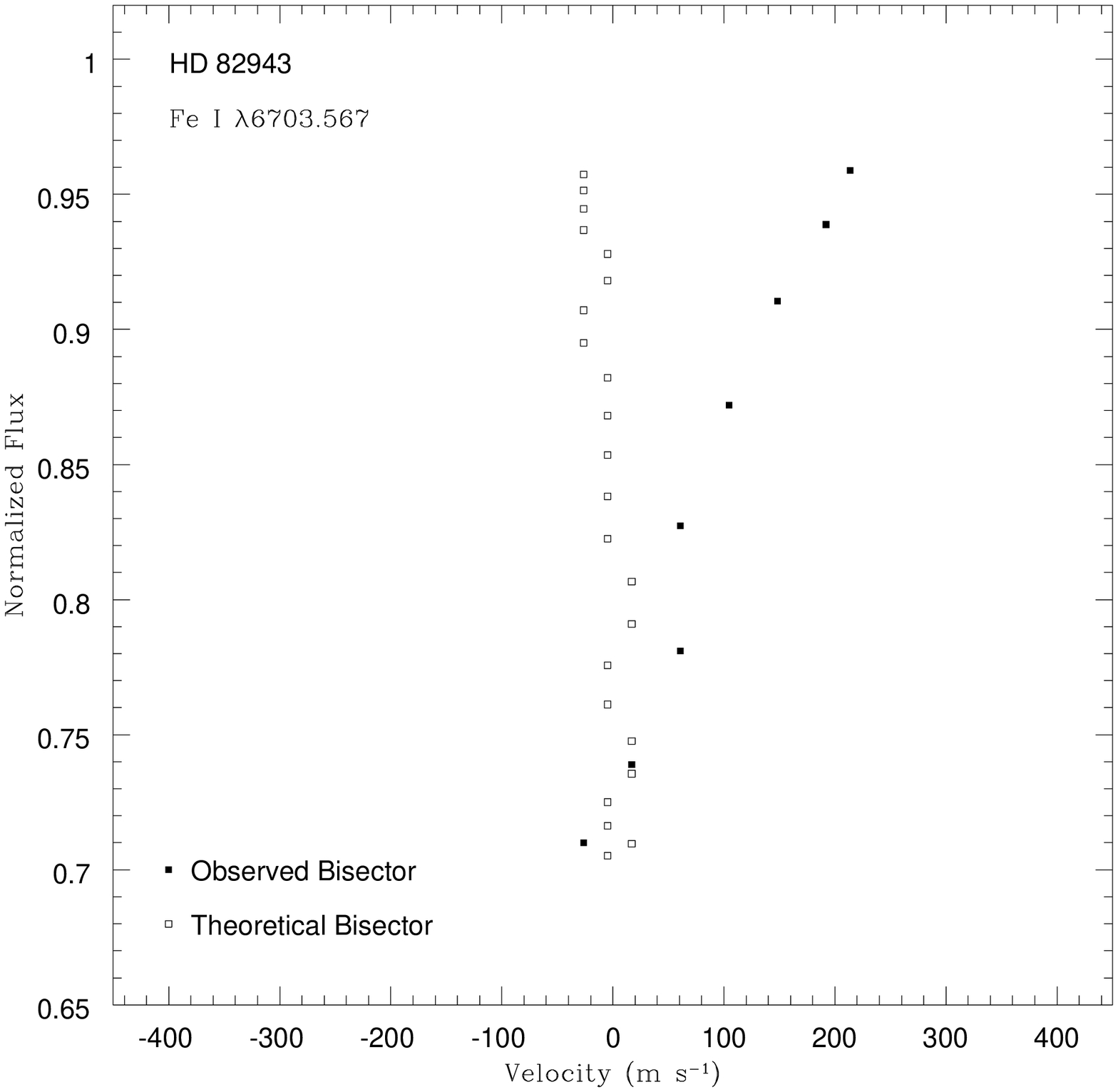}
\caption{Synthetic (open squares) and observed (filled squares) line bisectors 
measured for \ion{Fe}{1} at 6703.567 \AA. The asymmetry to the red in the observed
profile is indicative of photospheric velocity fields. Errors in the
bisectors are largest in the central portion of the line.}
\label{bisectors}
\end{figure}

Solar-type stars are known to display slight red asymmetries in spectral lines due to
the convective motions of granules (e.g. \citealt{ap02});
the flux across the line is dominated by hot, rising granules, while
the cooler, falling (i.e., red-shifted) inter-granule regions produce 
the small red asymmetries.  The line-bisector for the \ion{Fe}{1} 6703.567 \AA\ 
line is shown in Figure \ref{bisectors}, where it is plotted as
flux level versus velocity (in m s$^{-1}$) instead of wavelength. 
A detailed discussion of stellar line bisectors can be found in Gray (2005). 
The filled squares are the observed points in HD 82943, while the
open squares were derived from a synthesis of this region using
a standard 1D model, which necessarily produces a symmetric line profile.
The synthetic profile shows a vertical bisector, indicating an isolated
symmetric line (demonstrating that the \ion{Fe}{1} 6703.567 \AA\ line is good
for determining broadening parameters).
The observed line-bisector in HD 82943 deviates to the red due to convective
granules; the magnitude of the convective asymmetry
is manifested as a $\sim$ 200 m s$^{-1}$ excursion in the line bisector. 
The shape and amplitude of this bisector is typical for stars of
this type and represents a rather small perturbation of the line profile.

%%%%%%%%%%%%%%%%%%%%%%%%%%%%%%%%%%%%

%%%%%%%%%%%%%%%%%%%%%%%%%%%%%%%%%%%%

%%%%%%%%%%%%%%%%%%%%%%%%%%%%%%%%%%%

\subsection{The Line List}

\label{lines}

The determination of the $^{6}$Li/$^{7}$Li isotopic ratio rests upon
modeling the shape of the \ion{Li}{1} line-profile, which depends not only on the
isotopic ratios themselves, but also on the stellar broadening mechanisms
discussed in the previous section.  In addition, the shape of the \ion{Li}{1}
profile is affected by several weak absorption features that fall
within the wavelength region of the various lithium transitions.  There are 
several weak lines of CN and other metal absorption lines that will
blend with $^{6}$Li, thus a detailed compilation and careful assessment
of line lists is a critical ingredient for the analysis.

As mentioned in the introduction, the search for $^{6}$Li in planet hosting stars 
has been the subject of a few recent studies and line lists for the \ion{Li}{1} region are available
in the literature.  
As a starting point in this analysis, we adopted the line list from the most recent and 
complete study to date by \citet{m04}. The list of lines and atomic data (\textit{gf}-values and wavelengths)
carefully compiled by the authors were adjusted in that study to fit the \citet{k84} solar
spectrum. In addition, a laboratory carbon arc spectrum was used to adjust the CN lines (in both 
wavelengths and \textit{gf}-values) in comparison to theoretical line lists. 
As for the 3 possibilities for the unidentified feature at 6708.025 \AA\ listed by Mandell et 
al. (2004; \ion{Si}{1}, \ion{Ti}{1} and \ion{Ti}{2}) we adopted the \ion{Si}{1} line. Recall that Mandell et al. (2004)
found no significant differences in the $^{6}$Li/$^{7}$Li ratios derived using the 3 different lines.
The first step here was to return all atomic data in their line list to those values appearing in 
the original references. The CN molecular data were kept the same as in \citet{m04}.

This initial line list was checked via comparisons between the solar flux spectrum from \citet{k84}
and synthetic spectra which are based on the models and synthesis code adopted in this analysis.
The solar spectrum was modelled using an ATLAS9 ODFNEW model 
with $T_{\rm eff}$ = 5777 K, $\log g$ = 4.44, and  $\xi$ = 1.24 km $\rm s^{-1}$. 
The broadening parameters adopted were obtained from  fits to the 
unblended \ion{Fe}{1} line at 6703.567 \AA: $v$ sin \textit{i} = 1.70 km $\rm s^{-1}$, 
$V_{m}$ = 2.80 km $\rm s^{-1}$. The limb darkening coefficient,  $\epsilon$,
was taken from \citet{vm93}, however the 
choice of $\epsilon$ has little effect on the \ion{Li}{1} line
profile and no measurable impact on the estimation of $^{6}$Li isotopic fractions. 
The comparison of observed solar spectrum and synthesis in the small
wavelength interval around the \ion{Li}{1} feature was improved by a small
continuum adjustment of 0.25\%.  In addition, small
wavelength corrections to the nearby \ion{Fe}{1} line ($-$0.003 \AA) and to
\ion{Li}{1} (+0.002 \AA) line led to a better fit; these wavelength corrections
are those expected due to convective shifts in the solar
photosphere (see Section \ref{conv}).  A small change in the \ion{Fe}{1} \textit{gf}-value of +0.015 dex
improved the fit to the \ion{Fe}{1} profile, while the accurate laboratory
\textit{gf}-values for the \ion{Li}{1} components were maintained.  

\begin{figure}
\plotone{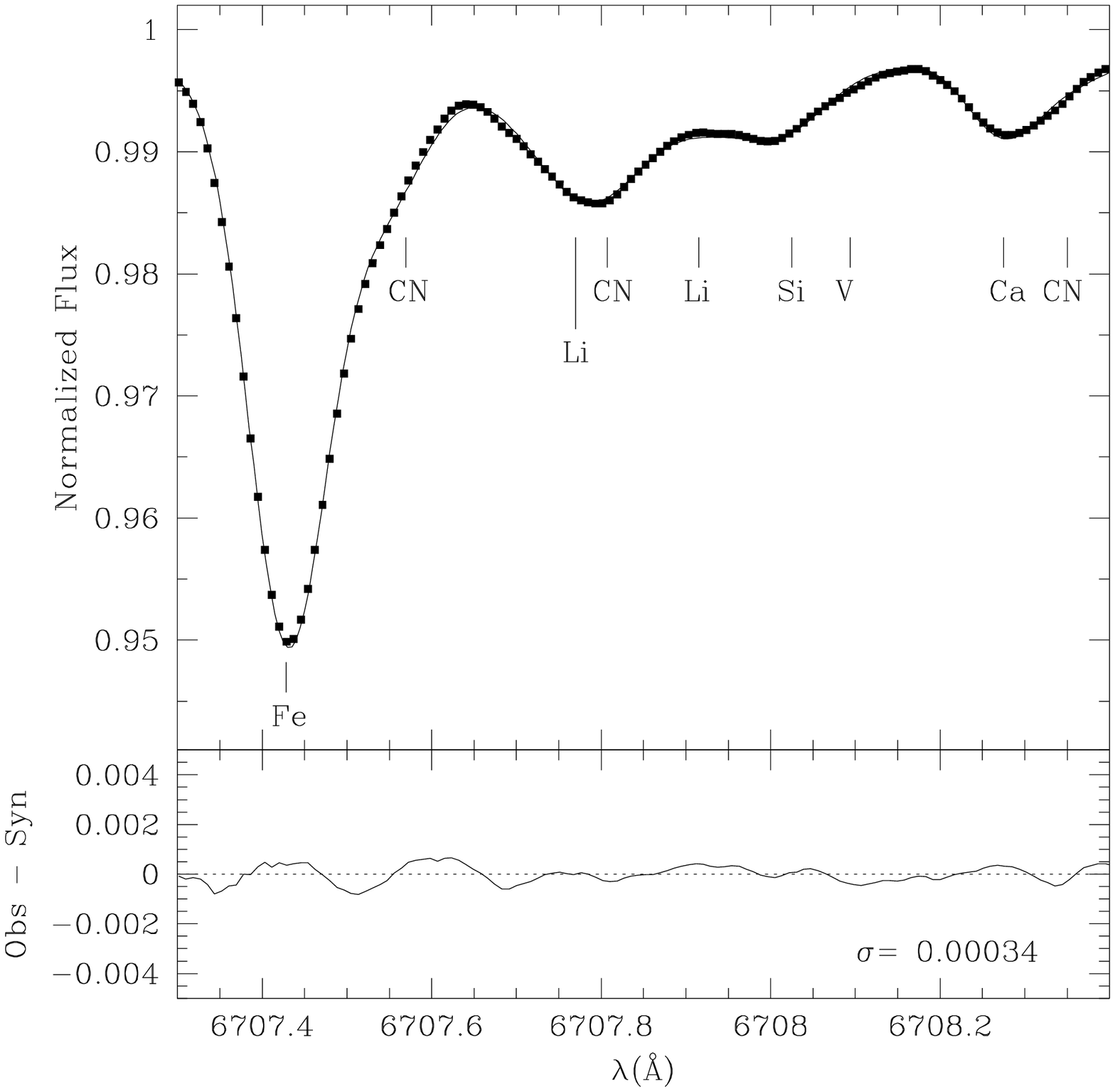}
\caption{Synthetic solar spectrum (solid line) computed with the line list in Table \ref{line_list}. The main
lines contributing to the synthesis are identified. The observed solar spectrum (filled squares) is from
\citet{k84}. The bottom panel shows the differences between model and observations; the overall
agreement is excellent.}
\label{solar_fit}
\end{figure}

Within the scope of the above changes, the fit to the solar spectrum
exhibited some small mismatch, especially near 6708.275 \AA, where there
appeared additional absorption.  
The absorption in this region was attributed to a \ion{Mg}{1} line by \citet{m04} 
and a \ion{V}{1} line by \citet{r02}.  No references to either of these 
lines could be found in a literature search, whereas \citet{k97} attributed
the absorption at this wavelength to a \ion{Ca}{1} line (present in VALD-2).  
There were other small differences between the line lists: in
the narrow region from 6708.31 to 6708.54 \AA\ the Mandell et al. list contained
4 CN lines while King et al. had 2 (at 6708.375 and 6708.635 \AA).
We included the \ion{Ca}{1} line plus the two adjacent CN lines;
some of the \textit{gf}-values were changed somewhat and wavelengths allowed to
shift by 0.001 \AA\ in order to improve the fit to the solar spectrum.
The final line list adopted in this study is presented in Table \ref{line_list}.
The best-fit to the solar spectrum achieved with this line list was excellent and 
this is shown in Figure \ref{solar_fit}. The best-fit was obtained for A(Li) = 0.96, 
which is very close to what was found by both \citet{k97} and \citet{r02},
and within 0.1 dex of the recommended value in \citet{a05}.

%%%%%%%%%%%%%%%%%%%%%%%%%%%%%%%%%%

\begin{deluxetable}{ccccc}
\tablecolumns{5}
\tablewidth{0pc}
\tablecaption{The Line List\label{line_list}}
\tablehead{
\colhead{$\lambda$} & \colhead{Identification} & \colhead{LEP} & \colhead{$\log gf$} & \colhead{Original} \\
\colhead{(\AA)} & \colhead{} & \colhead{(eV)} & \colhead{(dex)} & \colhead{Reference}}
\startdata
6706.548 & CN $Q_{2}$(93) (11,5)  & 3.130 & $-$1.359 & M04 \\
6706.567 & CN $Q_{2}$(80) (8,3)   & 2.190 & $-$1.650 & M04 \\
6706.657 & CN $R_{12}$(22) (7,3)  & 0.870 & $-$3.001 & M04 \\
6706.733 & CN $Q_{1}$(22) (7,3)   & 0.870 & $-$1.807 & M04 \\
6706.844 & CN $R_{1}$(34) (12,7)  & 1.960 & $-$2.775 & M04 \\
6706.863 & CN $P_{2}$(83) (7,2)   & 2.070 & $-$1.882 & M04 \\
6706.880 & \ion{Fe}{2}            & 5.956 & $-$4.504 & V   \\
6706.980 & \ion{Si}{1}            & 5.954 & $-$2.797 & V   \\
6707.205 & CN $Q_{2}$(47) (11,6)  & 1.970 & $-$1.222 & M04 \\
6707.282 & CN $Q_{2}$(60) (10,5)  & 2.040 & $-$1.333 & M04 \\
6707.371 & CN $Q_{1}$(85) (12,6)  & 3.050 & $-$0.522 & M04 \\
6707.431 & \ion{Fe}{1}            & 4.608 & $-$2.288 & R02 \\
6707.457 & CN $P_{12}$(13) (7,3)  & 0.790 & $-$3.055 & M04 \\
6707.470 & CN $Q_{1}$(28) (12,7)  & 1.880 & $-$1.451 & M04 \\
6707.473 & \ion{Sm}{2}            & 0.933 & $-$1.477 & V   \\
6707.518 & \ion{V}{1}             & 2.743 & $-$1.995 & V   \\
6707.545 & CN $Q_{2}$(44) (6,2)   & 0.960 & $-$1.548 & M04 \\
6707.595 & CN $Q_{2}$(29) (12,7)  & 1.890 & $-$1.851 & M04 \\
6707.596 & \ion{Cr}{1}            & 4.208 & $-$2.767 & V   \\
6707.645 & CN $P_{21}$(44) (6,2)  & 0.960 & $-$2.460 & M04 \\
6707.740 & \ion{Ce}{2}            & 0.500 & $-$3.810 & R02 \\
6707.752 & \ion{Sc}{1}            & 4.049 & $-$2.672 & V   \\
6707.756 & $^7$Li                 & 0.000 & $-$0.428 & H99 \\
6707.768 & $^7$Li                 & 0.000 & $-$0.206 & H99 \\
6707.771 & \ion{Ca}{1}            & 5.796 & $-$4.015 & R02 \\
6707.807 & CN $R_{1}$(64) (5,1)   & 1.210 & $-$1.853 & M04 \\
6707.848 & CN $R_{1}$(61) (19,12) & 3.600 & $-$2.417 & M04 \\
6707.899 & CN $P_{2}$(39) (20,13) & 3.360 & $-$3.110 & M04 \\
6707.907 & $^7$Li                 & 0.000 & $-$1.509 & H99 \\
6707.908 & $^7$Li                 & 0.000 & $-$0.807 & H99 \\
6707.919 & $^7$Li                 & 0.000 & $-$0.807 & H99 \\
6707.920 & $^6$Li                 & 0.000 & $-$0.479 & H99 \\
6707.920 & $^7$Li                 & 0.000 & $-$0.807 & H99 \\
6707.923 & $^6$Li                 & 0.000 & $-$0.178 & H99 \\
6707.930 & CN $Q_{21}$(35) (12,7) & 1.980 & $-$1.651 & M04 \\
6707.964 & \ion{Ti}{1}            & 1.879 & $-$6.903 & V   \\
6707.980 & CN $R_{21}$(72) (10,5) & 2.390 & $-$2.027 & M04 \\
6708.023 & \ion{Si}{1}            & 6.000 & $-$2.910 & I03 \\
6708.026 & CN $R_{2}$(35) (12,7)  & 1.980 & $-$2.031 & M04 \\
6708.073 & $^6$Li                 & 0.000 & $-$0.303 & H99 \\
6708.094 & \ion{V}{1}             & 1.218 & $-$3.113 & V   \\
6708.147 & CN $P_{2}$(42) (11,6)  & 1.870 & $-$1.434 & M04 \\
6708.275 & \ion{Ca}{1}            & 2.710 & $-$3.377 & K97 \\
6708.375 & CN                     & 1.979 & $-$1.097 & K97 \\
6708.499 & CN                     & 1.868 & $-$1.423 & K97 \\
6708.577 & \ion{Fe}{1}            & 5.446 & $-$2.728 & V   \\
6708.635 & CN $P_{2}$(42) (11,6)  & 1.870 & $-$1.584 & M04 \\
\enddata
\tablerefs{(H99) \citet{h99}; (I03) \citet{i03}; (K97) \citet{k97}; (M04) \citet{m04}; (R02) \citet{r02}; (V) VALD-2 \citep{k99}.}
\end{deluxetable}

%%%%%%%%%%%%%%%%%%%%%%%%%%%%%

%%%%%%%%%%%%%%%%%%%%%%%%%%%%%%%%%%%

%%%%%%%%%%%%%%%%%%%%%%%%%%%%%%%%%%%

\section{LITHIUM ABUNDANCES AND ISOTOPIC RATIOS}

\label{li}

As discussed previously,
several blending lines due to metals and CN affect the shape of the \ion{Li}{1} feature
in metal-rich stars.  It is possible to study which
of these potentially offending lines have the largest effects on the
$^{6}$Li absorption via test spectrum syntheses which can isolate each
of the nearby lines.

Synthetic spectra were generated to focus on lithium and each of the
possible blending species in turn: CN, \ion{Si}{1}, \ion{Ca}{1}, and \ion{V}{1}. In the case of
CN, the main impact on $^{6}$Li comes from 6707.807 \AA\ which falls slightly
blueward ($\sim$ 0.10 \AA) of most of the $^{6}$Li absorption and does
not affect the derived abundances significantly.  The \ion{Ca}{1} line at
6708.275 \AA\ falls too far to the red to affect $^{6}$Li, while \ion{V}{1}
at 6708.094 \AA\ is too weak to even be detectable.  The biggest effect
on $^{6}$Li absorption comes from the blending line due to \ion{Si}{1} at
6708.023 \AA, where its absorption falls on the red side of possible
$^{6}$Li absorption. Even here, absorption from $^{6}$Li that is
about 1\% deep would compete with \ion{Si}{1} (at solar metallicity) and
be detectable.  These tests indicate that $^{6}$Li absorption which is
$\sim$ 1\% deep would be marginally detected, while larger amounts of
$^{6}$Li that would produce absorption of only a few percent would be
detectable.

The important goal is to obtain an overall good fit across the entire
Li region from 6707.3 \AA\ to 6708.4 \AA\ and this includes fitting
the nearby feature just blueward of \ion{Li}{1} which is mainly \ion{Fe}{1} with some
contribution from CN.  Also, a close region nearly free of spectral lines
(6706.40 -- 6706.55 \AA) is useful for providing an estimate of the
local continuum level, while the wavelengths of \ion{Fe}{1} (6707.431 \AA) and
\ion{Li}{1}  were adjusted slightly to take into account convective shifts (see
Figure \ref{convective_shifts}). The abundances of Ca and V were allowed to vary within $\pm$0.2 dex
in order to improve the overall fits to the observed
spectra, as well as adjusting C and N to fit the blue CN feature.  The
final values, or limits, to $^{6}$Li were set by a $\chi_{r}^{2}$-minimization 
of A(Li), $^{6}$Li/$^{7}$Li and A(Si) simultaneously.

Lithium abundances and isotopic ratios are given in Table \ref{isotop} for the six
target stars. The uncertainties estimated for the Li abundances and
isotopic ratios were derived in a manner similar to that used for the
rotational and macroturbulent velocities.  The total lithium abundance
was varied (but keeping $^{6}$Li/$^{7}$Li fixed) in steps of 0.001 dex,
with the $\chi_{r}^{2}$ computed for each value. A change in $\chi_{r}^{2}$
of 1 was taken to define the $\pm$1$\sigma$-value of A(Li). The
uncertainty in the isotopic ratio was determined in an analogous way,
with $^{6}$Li/$^{7}$Li values varied in steps of 0.005 and the total
Li abundance being fixed. The errors in A(Li) and $^{6}$Li/$^{7}$Li
are also shown in Table \ref{isotop}.

In order to further test the sensitivity of our method to the isotopic ratios,
synthetic spectra were computed for 6 values of the $^{6}$Li/$^{7}$Li ratio
(between 0.00 to 0.05 with a step of 0.01) adopting the stellar parameters
derived for HD 82943. These model spectra were then sampled at the same
scale as bHROS, after which noise was added such that S/N = 750
(typical S/N for our observations). These \textquotedblleft degraded\textquotedblright\ 
synthetic spectra were
then subjected to the same analysis as the real spectra and we were able to
recover isotopic ratios within $\pm$0.01, demonstrating an analysis sensitivity
of 0.01 for $^{6}$Li/$^{7}$Li.

As an additional test, we used the line list in \citet{i03} in order to 
analyze the target star HD 82943. The line list adopted in that study is
identical to the one used by \citet{r02}, except for having a \ion{Si}{1} line
at 6708.025 \AA\ instead of a \ion{Ti}{1} line. All the input parameters (model atmospheres, 
broadening parameters and convective shifts for the \ion{Li}{1} and the nearby \ion{Fe}{1} line) 
and the analysis method employed were the same as described above;
the only difference being the removal of the Ca abundance as a free parameter 
because the feature at $\sim$ 6708.275 \AA\ is attributed to \ion{V}{1} in Israelian et al.'s list. 
The best fit between model and observations was achieved for A(Li) = 2.49 and $^{6}$Li/$^{7}$Li = 0.01,
which are in excellent agreement with the obtained results using the line list in Table \ref{line_list}. 
The derived isotopic ratio for this star, however, is significantly lower than the value 
derived in \citet{i03}; although the Li abundances in the two studies agree well. 

%%%%%%%%%%%%%%%%%%%%%%%%%%%%%%%%%%%

\begin{figure}
\plotone{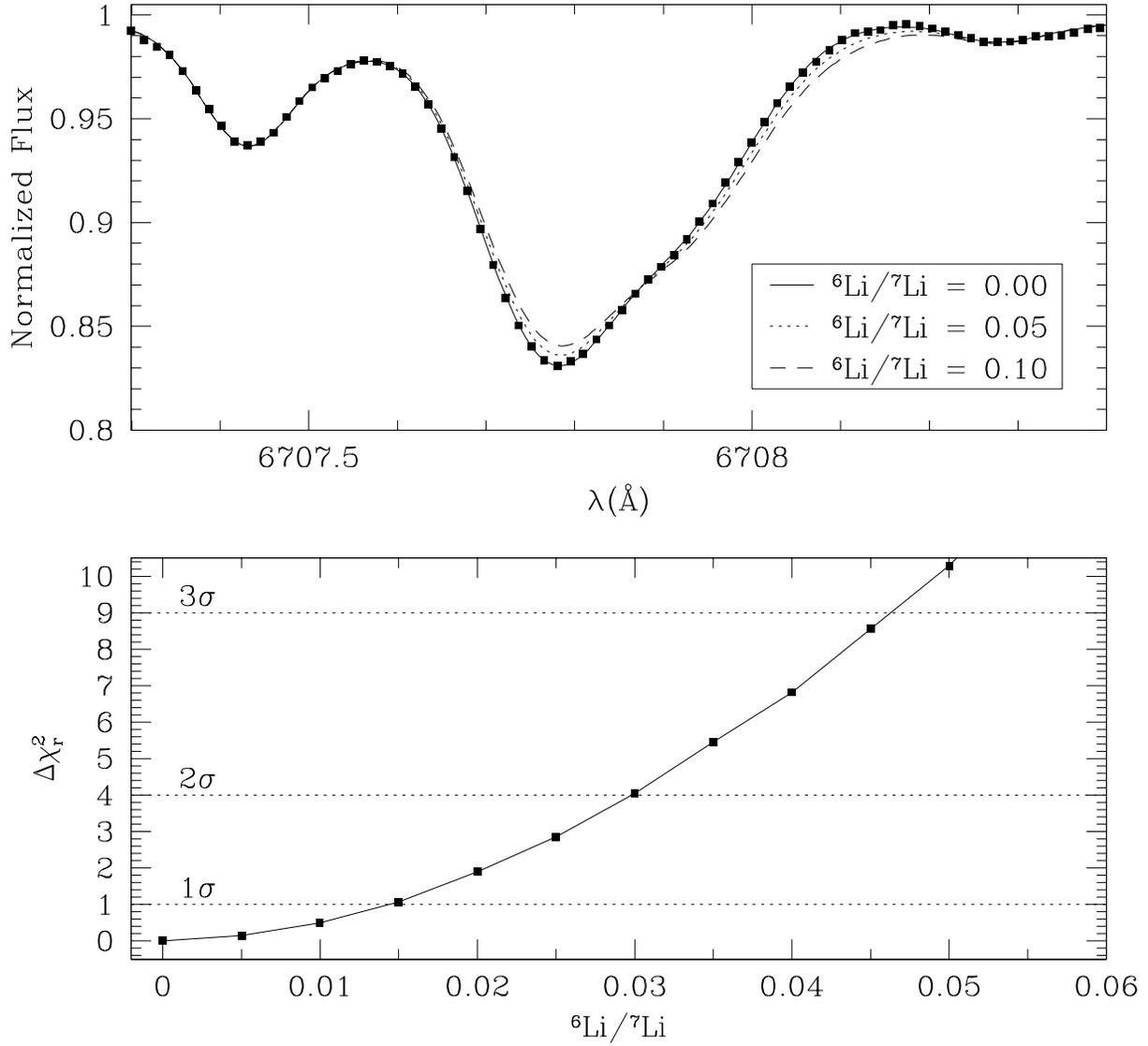}
\caption{Observed and synthetic \ion{Li}{1} profiles for HD 82943. The syntheses were calculated for
$^6$Li/$^7$Li isotopic ratios of 0, 5, and 10 \%. 
The Chi-squared minimization of the $^6$Li/$^7$Li isotopic ratio is 
shown in the bottom panel. The best fit is obtained for a synthetic spectrum with zero
contribution from $^{6}$Li.}
\label{hd82943_fit}
\end{figure}

%%%%%%%%%%%%%%%%%%%%%%%%%%%%%%%%%%%

%_________________________________________________________________

\section{DISCUSSION}

\label{disc}

\subsection{Three-Dimensional Model Atmospheres and Their Impact on the One-Dimensional 
$^{6}$Li Results}

\label{3D}

As discussed in Section \ref{conv}, convective granules result in small
red asymmetries in the spectral lines of solar-type stars, such as those
analyzed here for $^{6}$Li.  Since the $^{6}$Li components of the
neutral lithium ground-state doublet fall to the red of the $^{7}$Li
transitions and because $^{6}$Li is expected to be much less abundant
than $^{7}$Li, the presence of $^{6}$Li will result in a slight red
asymmetry to the combined \ion{Li}{1} feature.  Since both convection and 
$^{6}$Li can result in a red asymmetric profile, a discussion of the
expected convective asymmetries to the \ion{Li}{1} feature must be included
in order to interpret either detections or upper limits to $^{6}$Li
fractions.  

A more realistic theoretical treatment of the convective
motions that cause red-asymmetric line profiles has been included
in model stellar atmospheres via the incorporation of radiative
hydrodynamical convection, e.g. \citet{sn98} or \citet{a00}.
The resulting model atmospheric structures are three-dimensional,
time-dependent and contain self-consistent convective flows; such
models, often referred to as \textquotedblleft3D models\textquotedblright, naturally produce absorption
lines with small red asymmetries.  Recent analyses of the \ion{Li}{1} line profiles
in stars that span the T$_{\rm eff}$-range of the program stars here (although
mostly for much more metal-poor halo dwarfs and subgiants) have included a
discussion of effects due to 3D models (\citealt{a06}; Cayrel et al. 2007,2008).

\citet{a06} determined $^{6}$Li abundances for a set of 24 metal-poor 
halo dwarfs and subgiants based on very high quality spectra acquired with 
the ESO VLT/UVES. The authors found detections of $^{6}$Li at the 2$\sigma$ 
significance level in 9 stars.  They also showed that the isotopic ratio 
$^{6}$Li/$^{7}$Li is effectively immune 
to changes in the stellar parameters (at the levels of $\pm$100 K, $\pm$0.2 
dex, $\pm$0.2 dex, and $\pm$0.5 km/s in T$_{\rm eff}$, log g, [Fe/H], and 
microturbulent velocity, respectively). 
Another important result is that the usage of 3D model atmospheres in LTE 
(when compared to 1D models also in LTE) does not change significantly the
derived $^{6}$Li fractions from synthesis of the \ion{Li}{1} 6707.8 \AA\ line, 
suggesting that 1D models are essentially
as good as 3D models when employing LTE analyses.  
Finally, the authors observe that 1D $^{6}$Li/$^{7}$Li isotopic 
ratios are, in general, similar or lower than the ones derived from 3D LTE 
calculations. The average 3D--1D difference is +0.008 and the largest 
value is +0.033. In only 3 cases, are the 3D values smaller 
($-$0.010, $-$0.015, and $-$0.017) than the corresponding ones for the 1D case.
These results suggest that the upper limits for the isotopic ratios set here 
would probably increase 
if the analysis had been done with 3D.  However, the maximum increase would be 
compatible with the typical errors in the isotopic ratios.

More recently,  \citet{c07} showed that convection-induced line 
asymmetries could mimic the presence of $^{6}$Li, at the few percent level,
for the halo turn-off metal-poor star HD 74000. Note
that \citet{s98} found $^{6}$Li/$^{7}$Li = 0.00$\pm$0.02 for this star
via a 1D analysis, similar to the analysis performed in this study. 
The \citet{c07} result was based on a 3D hydrodynamical simulation for a 
single set of atmospheric parameters, while \citet{c08} presented 
theoretical line asymmetries for a range of atmospheric parameters that cover 
the values of the stars from \citet{a06}. Note that a
3D-NLTE treatment, the most adequate for the \ion{Li}{1} analysis according to 
\citet{a06}, is used by Cayrel (2007, 2008). The authors compute an 
asymmetry of about 2\%, which is exactly the mean value of $^{6}$Li/$^{7}$Li for
the stars from \citet{a06}.  Also, \citet{c08} investigate 
the dependence of these asymmetries on the stellar parameters 
T$_{\rm eff}$, log g, and [Fe/H].  Based on a small grid of 8 hydrodynamical 
simulations, they predict an increase in the \ion{Li}{1}
line asymmetry with metallicity for [Fe/H] $\ge$ $-$2.0.  Also, they find 
that hotter stars with lower gravities show larger asymmetries than 
cooler stars with higher gravities. 

Given the tight limits on $^{6}$Li fractions derived here via 1D measurements,
combined with the \citet{c08} predictions that there may be expected 
line asymmetries of the order of or larger than $\sim$2\% for more metal-rich
stars, there is no evidence of detectable $^{6}$Li in the 6 stars studied here.
The average value of $^{6}$Li/$^{7}$Li for the 6 stars in Table 5, inversily weighted
by their respective uncertainties, is $^{6}$Li/$^{7}$Li = 0.012, which is consistent
with zero $^{6}$Li within the errors. If the non-planet hosting star, HD 36435,
is rejected from the average due to its low S/N spectrum (Table \ref{obslog}), the weighted
average for the 5 stars with planets is even lower: $^{6}$Li/$^{7}$Li = 0.008.
With limits close to zero, these results also can provide tests on the 3D models
and NLTE analyses.

\subsection{Limits on Accreted Mass}

\label{mass}

An approximate upper limit to the number of $^{6}$Li atoms in the atmospheres
of the target stars can be estimated by combining limits to the number
of $^{6}$Li atoms with the convective zone masses.
To carry out this exercise we use the convective zone
masses for main sequence stars presented in \citet{p01}
in their Figure 1. This is a simple estimate which assigns a single
convective zone mass to the target stars (see table \ref{macc}) based on their effective
temperatures, but will provide a rough upper limit on the number of
$^{6}$Li atoms present in the outer layers of stellar atmospheres.  
The $^{6}$Li abundance is set by the $^{6}$Li/$^{7}$Li values in
Table \ref{isotop}; in those stars where this value is 0.00, the limit is set to
the 1$\sigma$ uncertainty.
Table \ref{macc} presents the resulting limits to the number of $^{6}$Li
atoms. In addition, the $^{6}$Li abundances are translated to upper limits of
accreted mass in Jovian masses. This is an admittedly
naive estimate, but does provide a framework in which to limit the
amount of accretion that could have occurred on the surfaces of these
5 planet-hosting stars.

%%%%%%%%%%%%%%%%%%%%%%%%%%%%%%%%%%%

\begin{deluxetable}{lccc}
\tabletypesize{\footnotesize}
\tablecolumns{4}
\tablewidth{0pc}
\tablecaption{Mass Accretion Limits\label{macc}}
\tablehead{
\colhead{Star} & \colhead{$\log$ M$_{\rm CZ}$} & \colhead{N} & \colhead{M$_{\rm acc}$} \\
\colhead{} & \colhead{(M$_{\odot}$)} & \colhead{($^{6}$Li)} & \colhead{(M$_{\rm Jup}$})} 
\startdata
HD 17051   & $-$2.44 & $<$4.6$\times10^{43}$ & $<$0.33 \\
HD 74156   & $-$2.23 & $<$1.0$\times10^{44}$ & $<$0.71 \\
HD 82943   & $-$2.15 & $<$6.2$\times10^{43}$ & $<$0.44 \\
HD 147513  & $-$1.90 & $<$3.8$\times10^{43}$ & $<$0.27 \\
HD 217107  & $-$1.66 & $<$2.9$\times10^{42}$ & $<$0.02 \\
\enddata
\end{deluxetable}

%%%%%%%%%%%%%%%%%%%%%%%%%%%%%%%%%%%

Accretion limits of a fraction of a Jovian mass are typical values derived
by both \citet{m04} and \citet{r02}. The tightest
accretion limit set here is for the coolest star, HD 217107, due to the
increasing strength of the \ion{Li}{1} line with decreasing T$_{\rm eff}$;
this star has a very low total lithium abundance with a corresponding
low limit of $^{6}$Li.  The
tighter limit on HD 217107 is perhaps also the most interesting from this
sample, as this star is the only one with a very closely orbiting
massive planet, with a sin \textit{i} = 0.073 AU and M$_{\rm planet}$ sin
\textit{i} = 1.33 M$_{\rm Jup}$.

While the non-detections of $^{6}$Li, at levels 1--2\% of total lithium,
are secure limits, the interpretation of this lack of $^{6}$Li in light
of accretion rests on details of stellar evolution. \citet{mr02}
have presented calculations of the expected evolution of
the $^{6}$Li abundance as a function of time after ingestion for various
stellar model masses and metallicities.  Within the standard model framework
of stellar evolution, where there is no exchange of material between the
convective envelope and the radiative interior, accreted $^{6}$Li would
survive for gigayears in main-sequence stars with M $\ge$ 1.0--1.1 M$_{\odot}$
and having metallicities of [Fe/H] = 0.0 and +0.3 \citep{mr02}. These types of stellar
masses encompass most of the target stars studied here, thus the limits
would suggest that any accretion of planetary material was typically less
than a few to several tenths of a Jovian mass.

When stellar evolution with non-standard transport processes, such as
microscopic diffusion or turbulent mixing, is used to investigate the fate
of $^{6}$Li on the surface of solar-type stars, the interpretation becomes
more complex, as discussed by \citet{mr02}.  Both diffusion
and turbulent mixing act to move material between the convective surface
layer and the deeper radiative interior, with the result that accreted
$^{6}$Li will be removed from the photosphere and destroyed.  The efficiency
of this destruction is a function of stellar mass, metallicity, and the
time at which $^{6}$Li is accreted.  For warmer effective temperatures,
T$_{\rm eff} \ge $ 6100 K, the survival of detectable quantities of $^{6}$Li
can exceed a Gyr, thus if accretion takes place after such stars settle onto
the main sequence, there would be measurable amounts of $^{6}$Li in some
of these stars. Large, sensitive surveys for $^{6}$Li remain useful ways
to probe and constrain accretion.

%_________________________________________________________________

\section{CONCLUSIONS}

\label{conc}

The main conclusion drawn from the analysis presented here is that no
detections of
$^{6}$Li are found in 5 planet-hosting stars which span the T$_{\rm eff}$
range of 5700--6100 K, masses from $\sim$1.1--1.4 M$_{\odot}$, and
metallicities from [Fe/H] $\sim$ +0.1--+0.4. Since the hotter, more
massive of the solar-type stars have the least massive convection zones,
the stars studied here would be some of the best candidates
for detecting signatures of accretion.  The combination of high
spectral-resolution with high-S/N makes this search one of the most
sensitive and the results here can be combined with previous studies
(\citealt{r02}; \citealt{m04}) to provide strong limits on
accretion of $^{6}$Li.

%_________________________________________________________________

\acknowledgements{We acknowledge the financial support of CNPq. We thank Jeremy King for discussions and for providing the line list for the spectral region between 6700 and 6720 \AA. 
KC thanks Martin Asplund for discussions.}

Based on observations obtained at the Gemini Observatory, which is operated by
the Association of Universities for Research in Astronomy, Inc., under a
cooperative agreement with the NSF on behalf of the Gemini partnership: the
National Science Foundation (United States), the Science and Technology
Facilities Council (United Kingdom), the National Research Council (Canada),
CONICYT (Chile), the Australian Research Council (Australia), Minist\'erio da
Ci\^encia e Tecnologia (Brazil) and Ministerio de Ciencia, Tecnolog\'ia e Innovaci\'on Productiva (Argentina).  Observations were obtained
in the following Gemini programs: GS-2006A-C-5 and GS-2006B-Q-47. Research here is supported
in-part by NASA GRANT NNH08AJ58I. 
Support for Simon C. Schuler has been provided by the NOAO Leo Goldberg Fellowship; NOAO is operated by the Association of Universities for Research Astronomy (AURA), Inc.,
under a cooperative agreement with the National Science Foundation.

{\it Facility:} \facility{Gemini:South(bHROS)}

%%%%%%%%%%%%%%%%%%%%%%%%%%%%%%%%%%


\begin{thebibliography}{}

 \bibitem[Allende Prieto et al.(2002)]{ap02} Allende Prieto, C., Asplund, M., Garc\'ia L\'opez, R. J., \& Lambert, D. L. 2002, \apj, 567, 544

 \bibitem[Anders \& Grevesse(1989)]{ag89} Anders, E., Grevesse, N. 1989, \gca, 53, 197

 \bibitem[Asplund et al.(2005)]{a05} Asplund, M., Grevesse, N., \& Sauval, A. J., in ASP Conf. Ser. 336, Cosmic Abundances as Records of Stellar Evolution and Nucleosynthesis, ed. F. N. Bash \& T. G. Barnes (San Francisco: ASP), 25

 \bibitem[Asplund et al.(2006)]{a06} Asplund, M., Lambert, D. L., Nissen, P. E., Primas, F., \& Smith, V. V. 2006, \apj, 644, 229

 \bibitem[Asplund et al.(2000)]{a00} Asplund, M., Nordlund, \AA., Trampedach, R., Allende Prieto, C., \& Stein, R. F. 2000, \aap, 359, 729

 \bibitem[Castelli \& Kurucz(2004)]{ck04} Castelli, F., \& Kurucz, R. L. 2004, Proceedings of the IAU Symp. No 210; IAU Symp. No 210, Modelling of Stellar Atmospheres, eds. N. Piskunov et al. 2003, poster A20 (astro-ph/0405087)

 \bibitem[Cayrel et al.(2008)]{c08} Cayrel, R., Steffen, M., Bonifacio, P., Ludwig, H.-G., \& Caffau, E. 2008, arXiv:0810.4290 [astro-ph]

 \bibitem[Cayrel et al.(2007)]{c07} Cayrel, R., Steffen, M., Chand, H., Bonifacio, P., Spite, M., Spite, F., Petitjean, P., Ludwig, H.-G., \& Caffau, E. 2007, \aap, 473, L37

 \bibitem[Ecuvillon et al.(2006)]{e06} Ecuvillon, A., Israelian, G., Santos, N. C., Mayor, M., \& Gilli, G. 2006, \aap, 449, 809

 \bibitem[Fischer \& Valenti(2005)]{fv05} Fischer, D. A., \& Valenti, J. 2005, \apj, 622, 1102

 \bibitem[Gonzalez(1997)]{g97} Gonzalez, G. 1997, \mnras, 285, 403

 \bibitem[Gonzalez(2006)]{g06} Gonzalez, G. 2006, \pasp, 118, 1494

 \bibitem[Gonzalez et al.(2001)]{g01} Gonzalez, G., Laws, C., Tyagi, S., \& Reddy, B. E. 2001, \aj, 121, 432

 \bibitem[Gray(2005]{g05} Gray, D. F. 2005, The Observation and Analysis of Stellar Photospheres (3rd ed.; Cambridge: Cambridge Univ. Press)

 \bibitem[Hinkle et al.(2000)]{h00} Hinkle, K., Wallace, L., Valenti, J., \& Harmer, D., ed. 2000, Visible and Near Infrared Atlas of the Arcturus Spectrum 3727-9300 \AA\ (San Francisco: ASP)

 \bibitem[Hobbs et al.(1999)]{h99} Hobbs, L. M., Thorburn, J. A., \& Rebull, L. M. 1999, \apj, 523, 797

 \bibitem[Israelian et al.(2001)]{i01} Israelian, G., Santos, N. C., Mayor, M., \& Rebolo, R. 2001, \nat, 411, 163

 \bibitem[Israelian et al.(2003)]{i03} Israelian, G., Santos, N. C., Mayor, M., \& Rebolo, R. 2003, \aap, 405, 753

 \bibitem[Israelian et al.(2004)]{i04} Israelian, G., Santos, N. C., Mayor, M., \& Rebolo, R. 2004, \aap, 414, 601

 \bibitem[King et al.(1997)]{k97} King, J. R., Deliyannis, C. P., Hiltgen, D. D., Stephens, A., Cunha, K., \& Boesgaard, A. M. 1997, \aj, 113, 1871

 \bibitem[Kurucz et al.(1984)]{k84} Kurucz, R. L., Furelind, I., Brault, J., \& Testerman, L. 1984, Solar Flux Atlas from 296 to 1300 nm (Cambridge: Harvard Univ. Press)

 \bibitem[Kupka et al.(1999)]{k99} Kupka, F., Piskunov, N., Ryabchikova, T A., Stempels, H. C., \& Weiss, W. W. 1999, \aaps, 138, 119

 \bibitem[Laws et al.(2003)]{l03} Laws, C., Gonzalez, G., Walker, K. M., Tyagi, S., Dodsworth, J., Snider, K., \& Suntzeff, N. B. 2003, \aj, 125, 2664

 \bibitem[Mandell et al.(2004)]{m04} Mandell, A. M., Ge, J., \& Murray, N. 2004, \aj, 127, 1147

 \bibitem[M\"{u}ller et al.(1975)]{m75} M\"{u}ller, E. A., Peytremann, E., \& de la Reza, R. 1975, \solphys, 41, 53

 \bibitem[Montalb\'an \& Rebolo(2002)]{mr02} Montalb\'an, J., \& Rebolo, R. 2002, \aap, 386, 1039

 \bibitem[Pasquini et al.(2007)]{p07} Pasquini, L., D\"{o}llinger, M. P., Weiss, A., Girardi, L., Chavero, C., Hatzes, A. P., da Silva, L., Setiawan, J. 2007, \aap, 473, 979

 \bibitem[Pinsonneault et al.(2001)]{p01} Pinsonneault, M. H., DePoy, D. L., \& Coffee, M. 2001, \apj, 556, L59

 \bibitem[Pollack et al.(1996)]{p96} Pollack, J. B., Hubickyj, O., Bodenheimer, P., Lissauer, J. J., Podolak, M., Greenzweig, Y. 1996, \icarus, 124, 62

 \bibitem[Ram\'irez \& Mel\'endez(2004)]{rm04} Ram\'irez, I., \& Mel\'endez, J. 2004, \apj, 609, 417

 \bibitem[Reddy et al.(2002)]{r02} Reddy, B. E., Lambert, D. L., Laws, C., Gonzalez, G., \& Covey, K. 2002, \mnras, 335, 100

 \bibitem[Santos et al.(2000)]{s00} Santos, N. C., Israelian, G., \& Mayor, M. 2000, \aap, 363, 228

 \bibitem[Santos et al.(2005)]{s05} Santos, N. C., Israelian, G., Mayor, M., Bento, J. P., Almeida, P. C., Sousa, S. G., \& Ecuvillon, A. 2005, \aap, 437, 1127

 \bibitem[Schuler et al.(2008)]{s08} Schuler, S. C., Margheim, S. J., Thirupathi, S., Asplund, M., Smith, V. V., Cunha, K., \& Beers, T. C. 2008, \aj, 136, 2244

 \bibitem[Smith et al.(1998)]{s98} Smith, V. V., Lambert, D. L., \& Nissen, P. E. 1998, \apj, 506, 405

 \bibitem[Sneden(1973)]{s73} Sneden, C. 1973, Ph.D. thesis, University of Texas, Austin

 \bibitem[Stein \& Nordlund(1998)]{sn98} Stein, R. F., \& Nordlund, \AA. 1998, \apj, 499, 914

 \bibitem[Th\'evenin(1990)]{t90} Th\'evenin, F. 1990, \aaps, 82, 179T

 \bibitem[Udry \& Santos(2007)]{us07} Udry, S., \& Santos, N. C. 2007, \araa, 45, 397

 \bibitem[Van Hamme(1993)]{vm93} Van Hamme, W. 1993, \aj, 106, 2096 

\end{thebibliography}
\end{document}